\documentclass[doublecol]{epl2}
\usepackage{epsfig}
\usepackage{amsmath}

\title{Efficient routing on multilayered communication networks}

\author{Jie Zhou\inst{1} \and Gang Yan\inst{1} \and Choy-Heng Lai\inst{1,2,3}}
\shortauthor{F. Author \etal}

\institute{
  \inst{1} Temasek Laboratories, National University of Singapore, Singapore
  117411\\
  \inst{2} Beijing-Hong Kong-Singapore Joint Centre for Nonlinear and Complex Systems (Singapore), National University of Singapore, Kent Ridge 119260,
  Singapore\\
  \inst{3} Department of Physics, National University of Singapore, Singapore 117542
}


\abstract{We study the optimal routing on multilayered
communication networks, which are composed of two layers of
subnetworks. One is a wireless network, and the other is a wired
network. We develop a simple recurrent algorithm to find an
optimal routing on this kind of multilayered network, where the
single-channel transmission mode and the multichannel transmission
mode used on the wireless subnetwork are considered, respectively.
Compared with the performance of the shortest path algorithm, our
algorithm can significantly enhance the transport capacity. We 
show that our methods proposed in this letter could take 
advantage of the coupling of the two layers to the most extent, 
so that the wireless subnetwork could sufficiently utilize the 
wired subnetwork for transportation.}

\begin{document}

\maketitle

\section{I. Introduction}

Transportation is an important problem in natural and engineering
systems. The study of transportation optimization on complex 
networks has attracted a lot of interest in the past 
decade~\cite{Newman:2003,Menezes:2004}. These studies include the 
influence of network structure on the 
transportation~\cite{Arenas:2001,Guimerra:2002,Toroczkai:2004,Tadic:2009,Xue:2010}, 
features of different transportation 
models~\cite{Echenique:2005,Zhao:2005,Cholvi:2005,Korniss:2006,Li:2010,Yang:2011}, 
and offer strategies to improve transportation from different 
viewpoints~\cite{Echenique:2004,Tang:2009}. The shortest path 
route, which takes the shortest-distance paths between the 
origins and the destinations, is a most common strategy for 
routing. However, this approach easily leads to congestion on 
some popular nodes having the most number of routes passing by. 
Therefore, to overcome this shortcoming several efficient 
strategies have been developed in a static scheme by focusing on 
the rerouting of congested flow so as to lead the system from 
congestion to free flowing~\cite{Yan:2006,Danila:2006}. It is 
proved that to find the optimal routing of a transportation 
network, which has maximum transportation capacity without 
congestion, is an \emph{NP}-hard problem. It is remarkable that 
Bassler \emph{et} \emph{al}~\cite{Danila:2006} recently presented 
routing strategies for wired networks and wireless networks 
respectively for the purpose of increasing the transportation 
capacity of these networks, which outperform many other methods. 
Their algorithm is a sub-optimal solution and takes the running 
time $O(N^3\log N)$. Hence, the transportation capacity can be 
much improved if one can judiciously choose the routes.

With the rapid development of communication networks, wired and
wireless networks have been widely applied in many aspects of
society, which include the Internet, which is based on the wired
infrastructure and mobile telecommunication, which is based on the
wireless framework. Despite the works introduced above, relatively
little attention has been paid to the integration of the wired and
wireless networks. The issue of taking the wired and wireless
networks as a whole subject has become increasingly fundamental
due to the widespread overlap of their applications. In this
paper, we introduce an intuitive model which captures the
important characteristics of the wired and wireless networks, and
then we study transportation optimization on such integrated 
networks. Specifically, we first construct a wired network and a 
wireless network. Then, we choose the same number of nodes in 
each network and combine them pairwise to form a new one. Thus, 
the previous wired and wireless networks become the two layers of 
the new network. Notice that some characteristics of the wireless 
network are different from the wired one. One is the absence of 
the effects of rich hub connectivity and small shortest path 
length, which by contrast are important features of wired 
networks. The other is a constraint in the process of packet 
transmission. The constraint comes from the broadcast 
transmission mode of the wireless device. When a wireless device 
is sending packets, other devices within the one-hop transmission 
range of the sender or the receiver are forbidden to send or 
receive information packets if they are working in the same 
channel, which is called medium access control. However, the 
wired networks do not have such constraint. After that, we 
propose a heuristic routing method to implement the efficient 
routine on the multilayered networks by considering the different 
features of the wired and wireless networks~\cite{Danila:2006}. 
Comparing with the shortest path method, we find our method can 
significantly enhance the transportation capacity of the 
two-layered network.

The remainder of this paper is structured as follows. In sec. II,
we present the model of the multilayered communication network. In
Sec. III, we present our algorithm for single channel transmission
mode and show the results of the algorithm. In Sec. IV, we study
the general situation where the wireless networks work in the
multichannel mode. In Sec. V, we give our conclusion.

\section{II. The Model}
\begin{figure}
\begin{center}
\epsfig{figure=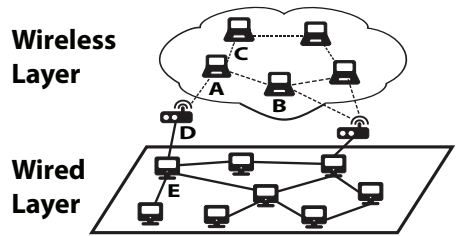,width=0.67\linewidth} 
\caption{Schematic illustration of the multilayered communication 
network with broadcast constraint, where node $A$, $B$ and $C$ 
are wireless nodes, node $D$ is an interfacing node, and node $E$ 
is an wired node. The dashed lines denote the wireless 
connections and the straight lines denote the wired connections. 
In this figure, the effective betweenness-to-capacity ratio of 
node $A$, denoted as $(B/C)_A^\mathrm{eff}$, is 
$(B/C)_A^\mathrm{eff}=(B/C)_A+(B/C)_B+(B/C)_C+(B/C)_D$, and 
correspondingly $(B/C)_D^\mathrm{eff}=(B/C)_D+(B/C)_A$ and 
$(B/C)_E^\mathrm{eff}=(B/C)_E.$ }\label{fig:illustration}
\end{center}
\end{figure}

We initially construct two subnetworks. One, called \emph{wired
layer}, is a wired network with $N_\mathrm{W}$ nodes and the
other, called \emph{wireless layer}, is a wireless network with
$N_\mathrm{L}$ nodes. The wired layer is constructed with the
Positive-Feedback Preference (PFP) Model~\cite{Zhou:2005} which 
accurately reproduces many topological properties of AS-level 
Internet. The wired layer constructed by the PFP model has the 
effects of rich hub connectivity and small shortest path length, 
and has an average degree $\simeq5.4$. For the wireless layer, we 
use the minimum degree geometric network 
model~\cite{Krause:2006}, which is appropriate in describing 
wireless networks. In the random geometric network model, a pair 
of nodes can make contact with each other when their distance is 
shorter than a critical value. However, this critical value needs 
to be provided by an external control authority which does not 
exist in \emph{ad hoc} networks. For the minimum degree geometric 
network model, all the nodes are randomly distributed on a square 
and they can decide on their own contact radius by increasing 
their contact power until they have at least $k_\mathrm{min}$ 
mutual neighbors. Therefore, with a given $k_\mathrm{min}$, all 
the nodal degrees in the wireless layer will be no less than 
$k_\mathrm{min}$. After generating the two layers, we randomly 
choose a node in the wired layer and a node in the wireless layer 
and then merge them into a single one. The newly merged node 
preserves all the connections that the previous two nodes have. 
Thus, this node could serve as an interface of the two layers. We 
continually perform this process until $N_\mathrm{I}$ pairs of 
nodes from the two layers are merged to form $N_\mathrm{I}$ 
nodes. For simplicity, we refer to these $N_\mathrm{I}$ nodes as 
\emph{interfacing node} which belong to the both layers, and we 
refer to the nodes in the \emph{wired} (\emph{wireless}) layer 
other than the interfacing nodes as \emph{wired} 
(\emph{wireless}) \emph{node}. Thus, $N_\mathrm{W}$ 
($N_\mathrm{L}$) is the number of nodes in the wired (wireless) 
layer and the size of the whole network is 
$N=N_\mathrm{W}+N_\mathrm{L}-N_\mathrm{I}$. 
Figure~\ref{fig:illustration} offers an illustration of the 
network structure. In this paper, we set the minimum degree 
$k_\mathrm{min}$ as $8$. We examined other values of 
$k_\mathrm{min}$ and find that the results are robust. In the 
simulations, all the averaged results and their standard 
deviation (which is indicated by the error bars), are obtained 
from $500$ different realizations, if not otherwise specified.


\section{III. Single Channel Mode}

Wireless networks have a broadcasting constraint when the nodes
work on the single channel mode~\cite{Krause:2006}. The constraint
is that when a node is sending an information packet, all its
incoming links are forbidden to send packets in order to avoid
packet collisions. This constraint causes a big obstruction for
the efficient information transmission. In the transmission
process, the wireless nodes work on the
first-in-first-possible-out basis. That is, a wireless node
attempts to send its first-in-line packet. When the intended
recipient is blocked due to the broadcasting constraint, the
node then tries to send its second-in-line packet and so on, until
either an idle recipient is found or the end of the queue is
reached. While, for the wired networks there is no such
constraint, and therefore the wired nodes simply work on the
first-in-first-out basis. The interfacing nodes adopt the
transmission process according to whether the intended recipient
or the sender is a wired node or a wireless node.

When a new packet is added to the network at a node, it is
appended at the end of the queue of this node. A packet is removed when it
reaches its destination. When the routes between all pair of nodes
are given, the betweenness of a node $i$ is defined as
$B_i=\Sigma_{s\neq t}p_i(st)/p(st)$, where $p(st)$ is the number
of routes going from node $s$ to node $t$ and $p_i(st)$ is the
number of routes going from node $s$ to node $t$ and passing
through node $i$. We assume that a node $i$ has the processing
capacity of handling $C_i$ information packets per time step and
different nodes may have different processing capacities. Thus,
the value of the betweenness-to-capacity ratio of node $i$,
denoted as $(B/C)_i$, equals to $B_i/C_i$.

An important consequence of the broadcast constraint is that a
node in the wireless layer needs not only to process the
information packets on itself but also to process the packets on
its wireless neighbors. This situation leads to a modified version
of the betweenness-to-capacity ratio~\cite{Krause:2006}, which is
the summation of the betweenness-to-capacity ratios of the
objective node and its wireless neighbors. However, the wired
nodes do not need such modification. Hence, the effective
betweenness-to-capacity ratio, denoted as $(B/C)^\mathrm{eff}$,
for different nodes is different. The effective
betweenness-to-capacity ratio of a wireless node is equal to the
sum of its own betweenness-to-capacity ratio and the
betweennesses-to-capacity ratio of all its neighbors. The
effective betweenness-to-capacity ratio of an interfacing node is
equal to the sum of its own betweenness-to-capacity ratio and the
betweennesses-to-capacity ratio of its neighbors connected with 
its wireless connections. Finally, the effective 
betweenness-to-capacity ratio of a wired node is equal to its own 
betweenness-to-capacity ratio (see Fig.~\ref{fig:illustration}). 
The maximum transportation capacity depends on the maximum 
effective betweenness-to-capacity ratio which is denoted as 
$(B/C)_\mathrm{max}^\mathrm{eff}$. The optimal routing is 
achieved when $(B/C)_\mathrm{max}^\mathrm{eff}$ is minimized. 
Adopting the algorithm that is used for single layered 
network~\cite{Danila:2006}, we extend it to the multilayered 
communication network to reduce the 
$(B/C)_\mathrm{max}^\mathrm{eff}$ through the following steps:

1. Assign every link a unit weight and compute the shortest path
between all pairs of nodes and the betweenness of every node.

2. Calculate the effective betweenness-to-capacity ratio of all
the nodes and find the node which has the highest score. (a) If
the node is on the wireless layer, increase the weights of all the
links sourced from this node and its neighbors by half unit
weight. (b) If the node is an interfacing node, increase the
weights of the links sourced from this node and its neighbors in
the wireless layer by half unit weight. (c) If the node is on the
wired layer, increase the weight of the links sourced from this
node by half unit weight.

3. Recompute the shortest paths and the betweenesses. Go back to
step 2.

4. The algorithm stops when the value of the maximum effective
betweenness-to-capacity ratio is stable.

We note that the weight of a link can be regarded as the measure 
of the length of this link. Hence, a larger weight corresponds to 
a longer distance, and a packet will less likely pass through the 
corresponding link because of the higher cost required.

\begin{figure}
\epsfig{figure=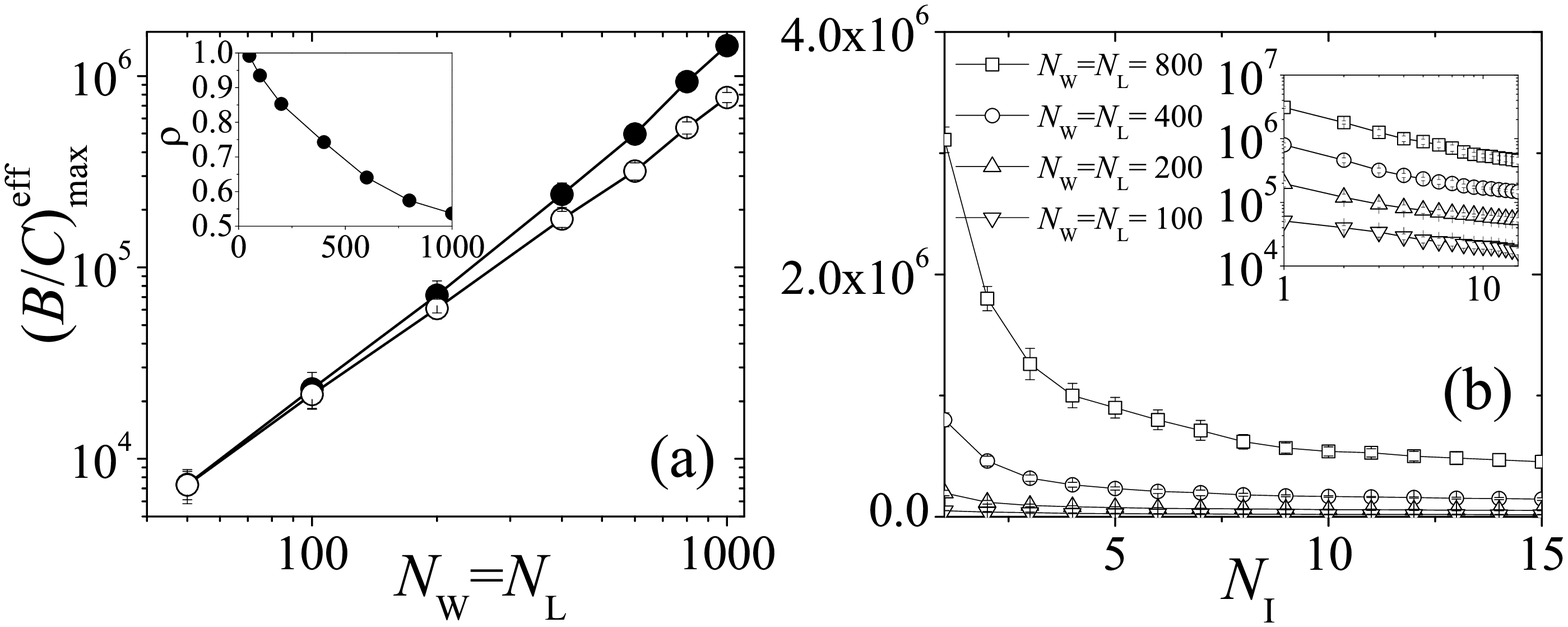,width=1\linewidth} \caption{(a) The 
maximum effective betweenness-to-capacity ratio for the shortest 
path algorithm (solid symbols) and for the optimal algorithm 
(open symbols) as a function of $N_\mathrm{W}$ and 
$N_\mathrm{L}$. In each network, $N_\mathrm{I}=10$ and 
$N_\mathrm{W}=N_\mathrm{L}$. The inset shows the ratio $\rho$ of 
the results obtained from the optimal algorithm to those obtained 
from the shortest algorithm. (b) The value of 
$(B/C)_\mathrm{max}^\mathrm{eff}$ after performing the optimal 
algorithm as a function of the number of interfacing nodes for 
different network sizes. The inset is a log-log plot of the main 
plot. In this figure, all the nodes have the capacity 
$C=1$.}\label{fig:Single}
\end{figure}

A plot of the maximum effective betweenness-to-capacity ratio 
$(B/C)_\mathrm{max}^\mathrm{eff}$ obtained from the shortest path 
method and the optimal algorithm versus network size is shown in 
Fig.~\ref{fig:Single}(a). Obviously, 
$(B/C)_\mathrm{max}^\mathrm{eff}$ increases with the increasing 
of the size of the networks. The inset shows the ratio $\rho$ of 
the results obtained from the optimal algorithm to those obtained 
from the shortest path algorithm. We observe that when the 
network size is small, the effect of the optimal algorithm is 
very limited. This is because of the fact that those small 
networks do not leave much space for the algorithm to adjust the 
transmission routes to reduce the 
$(B/C)_\mathrm{max}^\mathrm{eff}$. However, with the growth of 
the network size, there could be more alternative routes for a 
transmission. Therefore, it could be easier for the algorithm to 
find a route without passing the node that has the maximum 
betweenness-to-capacity  ratio, which results in a considerable 
decrease in the $(B/C)_\mathrm{max}^\mathrm{eff}$. Hence, this 
algorithm could be more effective for a larger network, which is 
a good point as real communication networks are generally very 
large.

The interfacing nodes take on an important role for the whole
network. Since the routes of the inter-layer communication need 
to pass through the interfacing nodes, when the number of 
interfacing nodes is small they may be the bottleneck. 
Figure~\ref{fig:Single}(b) shows the maximum effective 
betweenness-to-capacity ratio $(B/C)_\mathrm{max}^\mathrm{eff}$ 
from the optimal algorithm as a function of $N_\mathrm{I}$ for 
different network sizes. We can see that when $N_\mathrm{I}$ is 
small the value of $(B/C)_\mathrm{max}^\mathrm{eff}$ can be very 
large, which means the algorithm has limited effect in this 
situation because of the bottleneck effect of the interfacing 
nodes. However, with the increase of the $N_\mathrm{I}$ the value 
reduces quickly and the situation gets much improved.

\section{IV. Multichannel Mode}
Single channel mode is the primary mode for wireless
communication, but its efficiency is low. Thus, this mode is not
suitable for large networks. In fact, the multichannel mode is
widely adopted in engineering, and explicit regulations have been
developed for proper operation~\cite{IEEE802.11:2007}. A common
way is to let wireless devices consistently change their working
channel according to a pre-encoded random sequence. In this way,
when a group of nodes located nearby are transmitting information,
the possibility of them using the same channel is low. Even when
this happens they can simply turn to other channels to avoid such
situation. Thus, the situation in which different pairs of 
communicating nodes happen to work on the same channel is well 
avoided. Since nodes that work on different channels do not 
influence each other, the effect of the broadcast constraint can 
be ignored in the multichannel mode and the communication 
efficiency could be significantly enhanced. In the following, we 
study the transport optimization of the multilayered network in 
which the wireless layer works in the multichannel mode.

Since the broadcasting constraint can be ignored under the
multichannel mode, it is not necessary to use the modified version
of betweenness-to-capacity ratio as did in the previous section. 
In this mode, the effective betweenness-to-capacity ratio is just 
the betweenness-to-capacity ratio of the node itself. Thus, in 
the multichannel mode the algorithm works as follows:

1. Assign every link a unit weight and compute the shortest path
between all pairs of nodes and the betweenness of every node.

2. Calculate the betweenness-to-capacity ratio of all the nodes
and find the node which has the highest value. Increase the weight
of the links reached by this node by half unit weight.

3. Recompute the shortest paths and the betweennesses. Go back to
step 2.

4. The algorithm stops when the value of the maximum
betweenness-to-capacity ratio is stable.

\begin{figure}
\epsfig{figure=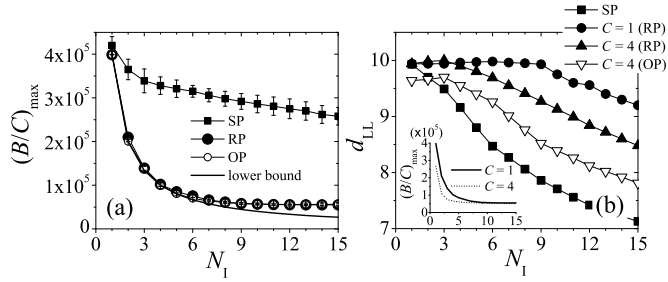,width=1\linewidth} \caption{ (a) 
$(B/C)_\mathrm{max}$ as a function of $N_\mathrm{I}$ for the 
shortest path algorithm (black square symbols), the optimal 
algorithm for Random Placement (RP) (solid circle symbols), the 
optimal algorithm for Optimal Placement (OP) (open circle 
symbols) and the theoretical lower bound which is 
$(2N_\mathrm{W}N_\mathrm{L})/(N_\mathrm{I}C)$.  (b)  Average 
transmission distance between wireless nodes $d_\mathrm{LL}$ as a 
function of $N_\mathrm{I}$ for the shortest path algorithm (solid 
square symbols), the optimal algorithm for RP with $C=1$  (solid 
circle symbols), the optimal algorithm for RP with interfacing 
nodes having $C=4$ (solid triangle symbols), and the optimal 
algorithm for OP with interfacing nodes having $C=4$ (open 
triangle symbols). The inset shows the $(B/C)_\mathrm{max}$ of 
the cases that the interfacing nodes have $C=1$ and $C=4$ as a 
reference. All the networks in this figure have 
$N_\mathrm{W}=200$ and $N_\mathrm{L}=1000$. All the nodes other 
than the interfacing nodes have $C=1$.}\label{fig:Multiple}
\end{figure}

An interesting point of the multilayered network is the influence 
of the interfacing nodes. Since all the inter-layered 
transmission need to go through the interfacing nodes, we may 
estimate the lower bound of the maximum betweenness of the 
interfacing nodes. From the definition of the betweenness, we 
have $\sum_{i\in\mathcal{I}}B_i=\sum_{i\in\mathcal{I}}\sum_{s\neq 
t}p_i(st)/p(st)$, where $\mathcal{I}$ is the set of the 
interfacing nodes. This equation could be further separated into 
four cases. Those are (i) $s\in\mathcal{W}$, $t\in\mathcal{W}$; 
(ii) $s\in\mathcal{L}$, $t\in\mathcal{L}$; (iii) 
$s\in\mathcal{W}$, $t\in\mathcal{L}$; and (iv) $s\in\mathcal{L}$, 
$t\in\mathcal{W}$; where $\mathcal{W}$ and $\mathcal{L}$ are the 
sets of the wired nodes and the wireless nodes, respectively. 
Since each packet for inter-layered communication passes through 
the interfacing nodes at least once, for cases (iii) and (iv) we 
have $\sum_{i\in\mathcal{I}}\sum_{s\in\mathcal{W}, 
t\in\mathcal{L}}p_i(st)/p(st)\geq N_\mathrm{W}N_\mathrm{L}$ and 
$\sum_{i\in\mathcal{I}}\sum_{s\in\mathcal{L}, 
t\in\mathcal{W}}p_i(st)/p(st)\geq N_\mathrm{W}N_\mathrm{L}$. The 
equal sign is obtained when each packet for inter-layered 
communication passes through the interfacing nodes only once. 
Therefore, even all the routes for intra-layered communication 
are restricted in the respective layer, that is 
$p_{i\in\mathcal{I}}(st)=0$ for cases (i) and (ii), we have 
$\sum_{i\in\mathcal{I}}B_i\geq 2N_\mathrm{W}N_\mathrm{L}$. Since 
there are $N_\mathrm{I}$ interfacing nodes, the maximum 
betweenness of the interfacing nodes should be no less than 
$2N_\mathrm{W}N_\mathrm{L}/N_\mathrm{I}$. This value is reached 
when the interfacing nodes take the same amount of the 
information packets transmitted between the layers and routes of 
the intra-layered communication are confined in that layer. Thus, 
when $N_\mathrm{I}$ is small, the largest betweenness of the 
interfacing nodes can be very large. 
Figure~\ref{fig:Multiple}\,(a) shows the maximum 
betweenness-to-capacity ratio $(B/C)_\mathrm{max}$ as a function 
of $N_\mathrm{I}$ obtained from the shortest path algorithm 
(square symbols), the optimal algorithm (circle symbols), and the 
theoretical lower bound of the interfacing nodes 
$(2N_\mathrm{W}N_\mathrm{L})/(N_\mathrm{I}C)$ (black curve), for 
the case of $C=1$, $N_\mathrm{W}=200$ and $N_\mathrm{L}=1000$. 
Similarly to the single channel mode, $(B/C)_\mathrm{max}$ 
decreases with the increase of $N_\mathrm{I}$, which means there 
also exists the bottleneck effect in the multichannel mode when 
the number of interfacing nodes is small. Moreover, the circle 
symbols almost overlap with the black curve when 
$N_\mathrm{I}\leq 6$ and are just a little larger when 
$N_\mathrm{I}>6$, which shows that the optimal algorithm can 
reduce the $(B/C)_\mathrm{max}$ nearly to the theoretical lower 
bound and thus to the optimal level. We note that when 
$N_\mathrm{I}=1$ the values of $(B/C)_\mathrm{max}$ of the two 
algorithms are very close. This is because when there is only one 
interfacing node, this node has to handle all the inter-layer 
communications whether or not the optimal algorithm is adopted, 
which causes the node to have the $(B/C)_\mathrm{max}$ and the 
value is approximated to 
$(2N_\mathrm{W}N_\mathrm{L})/(N_\mathrm{I}C)=4\times10^5$. 
Furthermore, for the optimal algorithm when $N_\mathrm{I}=2$ the 
value of $(B/C)_\mathrm{max}$ drops by     around a half compared 
to the case of $N_\mathrm{I}=1$, which means this algorithm can 
well distribute the information packets to the two interfacing 
nodes. Furthermore, we observe that when $N_\mathrm{I}$ is small 
the distribution of $(B/C)_\mathrm{max}$ obtained from the 
optimal algorithm under different realizations is squeezed in a 
narrow range. For example, for $N_\mathrm{I}=1$, the ratio of the 
standard deviation of the distribution of $(B/C)_\mathrm{max}$ 
with performing the algorithm to that without performing the 
algorithm is smaller than $0.01$. This phenomenon is distinct 
from what is observed in the single channel mode. In the single 
channel mode, the distributions of the 
$(B/C)_\mathrm{max}^\mathrm{eff}$ with and without performing the 
optimal algorithm spread in the ranges with comparable sizes, and 
both can be well fitted by the Gumbel distribution 
\cite{Moreira:2002}. The reason of this phenomenon is that in the 
multichannel mode, the optimal algorithm could always tune the 
$(B/C)_\mathrm{max}$ close to the theoretical lower bound 
resulting in the fact that the values of $(B/C)_\mathrm{max}$ for 
different realizations are very near to each other, which further 
proves the effectiveness of the algorithm.

As the maximum betweenness-to-capacity ratio of the interfacing 
nodes should be no less than 
$(2N_\mathrm{W}N_\mathrm{L})/(N_\mathrm{I}C)$, when the value of 
$(B/C)_\mathrm{max}$ is close to this lower bound, the routes for 
the intra-layered communication need to be mainly restricted in 
the respective layer to avoid going through the interfacing 
nodes. Thus, two wireless nodes cannot use the wired layer as 
well as its effects of rich hub connectivity and small shortest 
path length to shorten their distance in this situation. In order 
to improve the transmission efficiency by reducing the 
transmission distance, without increasing the value of 
$(B/C)_\mathrm{max}$, an effective method is to increase the 
processing capacity of the interfacing nodes. To illustrate the 
effectiveness of this method we show the average transmission 
distance among all pairs of wireless nodes, denoted as 
$d_\mathrm{LL}$, in Fig.~\ref{fig:Multiple}\,(b). In this figure, 
$d_\mathrm{LL}$ obtained from the shortest path algorithm as 
shown by the solid square symbols is the smallest one among all 
the cases, while $d_\mathrm{LL}$ obtained from the optimal 
algorithm where all the nodes have $C=1$ as shown by the solid 
circle symbols is much greater especially for large 
$N_\mathrm{I}$. However, when we increase the processing capacity 
of the interfacing nodes to $C=4$ while other nodes still have 
$C=1$, the $d_\mathrm{LL}$ has an obvious drop as shown by the 
solid triangle symbols which shows that increasing the processing 
capacity of the interfacing nodes could be of benefit to the 
wireless nodes utilizing the wired layer to reduce the 
transmission distance.

Moreover, we consider that if the positions of the interfacing 
nodes in the wireless layer are well-distributed, the wireless 
nodes could access their nearest interfacing nodes with the least 
of hops on average. Thus, this placement could further shorten 
the transmission distance when wireless nodes utilize the wired 
layer for transmission. For simplicity, we call such placement of 
the interfacing nodes in the wireless layer as \emph{Optimal 
Placement} (denoted as ``OP"), and correspondingly the placement 
in which the interfacing nodes are randomly distributed in the 
wireless layer as we did in the previous part of the work is 
called \emph{Random Placement} (denoted as ``RP"). To obtain the 
OP, we first get $N_\mathrm{I}$ optimal positions with the 
algorithm introduced in the appendix. Then, in the stage of 
constructing multilayered network we choose $N_\mathrm{I}$  
wireless nodes who are nearest to the $N_\mathrm{I}$ optimal 
positions as the wireless candidates to compose the 
$N_\mathrm{I}$ interfacing nodes. In this way, the composed 
interfacing nodes have OP in the wireless layer. The open 
triangle symbols in Fig.~\ref{fig:Multiple}\,(b) show the results 
of the OP where the interfacing nodes have $C=4$. We observe that 
the $d_\mathrm{LL}$ is further reduced in this case. The inset 
shows the corresponding $(B/C)_\mathrm{max}$ for the cases that 
the interfacing nodes have $C=1$ and $C=4$. Obviously, the 
$(B/C)_\mathrm{max}$ for the case of $C=4$ is always smaller than 
that of $C=1$. We note that $(B/C)_\mathrm{max}$ of OP and RP are 
very close when the capacity of the interfacing nodes in both 
cases is the same. An example is shown in 
Fig.~\ref{fig:Multiple}(a) where the open circle symbols 
correspond to OP and solid circle symbols correspond to RP. 
Therefore, by combining the methods of using the optimal 
algorithm, increasing the capacity of the interfacing nodes which 
are a small fraction of the whole network, and the OP, both the 
transportation capacity and average transmission distance can be 
much improved. Since these two factors are generally conflicting 
with each other, it is significant that both of them can be 
enhanced at the same time.

\section{V. Conclusion}

In summary, we propose a model to describe multilayered
communication networks. In the model, there are two layers. One,
called wired layer, is composed of nodes connected with wired 
links, and the other, called wireless layer, is composed of nodes 
connected with wireless links. The layers are connected by 
merging some pairs of nodes in different layers into single ones, 
called \emph{interfacing nodes}. Then, we present a recurrent 
algorithm to find optimal routing for this multilayered network 
for two different cases in which the wireless nodes can work in 
single channel mode and multichannel mode, respectively. In this 
algorithm, by gradually increasing the weight of the connections 
pointing to the node which has the maximum effective 
betweenness-to-capacity ratio, the burden on this node could be 
gradually reduced. By repeating this process, the final maximum 
effective betweenness-to-capacity ratio could be much suppressed 
and the transportation capacity could be enhanced significantly. 
Our results show that this algorithm is an effective navigation 
technique for multilayered communication networks. Since for the 
network on one hand the wireless layer may utilize the wired 
layer for transportation, on the other hand the interfacing nodes 
can easily become bottleneck, how to utilize the advantage of the 
multilayered networks while avoiding the restriction of the 
interfacing nodes is crucial for enhancing the transportation 
capacity. Our routing method proposed for the multilayered 
networks can squeeze the interfacing nodes to the extent that 
their effective betweenness-to-capacity ratio is almost the same, 
so that the bottleneck effect is suppressed to the most extent. 
By further cooperating with the methods of increasing the 
processing capacity and the OP of the interfacing nodes, the 
wireless layer may have a sufficient utilization of the wired 
layer for transportation. Hence, our method could serve as a good 
candidate of the standard technique for optimal transportation on 
multilayered communication networks.

\section{Appendix. A method of placing an arbitrary number of nodes on a square in a well-distributed pattern}

\begin{figure}\begin{center}
\epsfig{figure=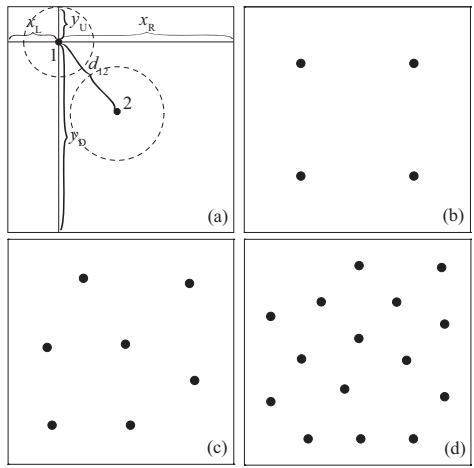,width=0.7\linewidth}\end{center} 
\caption{ (a) An example of $N=2$, where labels $1$ and $2$ 
indicate the centers of circle $1$ and $2$. We denote the 
candidate radius of circle $1$ and $2$ as $r_1$ and $r_2$, 
respectively. For circle $1$, $x_\mathrm{L}$, $x_\mathrm{R}$, 
$y_\mathrm{U}$ and $y_\mathrm{D}$ are the distances between its 
center to the four boundaries of the square, and $d_{12}$ is the 
distance between the two centers. Hence, 
$r_1=\mathrm{min}\{x_\mathrm{L},x_\mathrm{R},y_\mathrm{U},y_\mathrm{D},d_{12}/2\}$. 
Similarly, we can get the value of $r_2$. Thus, the value of 
$r_\mathrm{min}$ is $r_\mathrm{min}=\mathrm{min}\{r_1, r_2\}$. 
Dashed circles are those plotted by the candidate radii. (b), (c) 
and (d) show the examples of the cases $N=4$, $7$ and $15$, 
respectively.}\label{fig:OptimalPlacement}
\end{figure}

Suppose there are $N$ nodes to be dealt with, we first regard the 
$N$ nodes as $N$ solid circles with the same radius which are 
required to be confined in the square. It can be proved that 
placing the nodes on the square so that they are well-distributed 
is equivalent to place these solid circles so that they can have 
the largest identical radius, and the positions of the centers of 
these circles are the positions in which to place the nodes. 
Following we present the algorithm of placing these solid circles 
so that they can have the largest identical radius:

1. Initially, randomly distribute these nodes on the square. 
Then, assign each node a candidate radius which is the shortest 
one among the distances that are from the node to the four 
boundaries of the square or as half as those from the node to the 
other nodes. The identical radius of these solid circles which 
satisfies the above conditions, denoted as $r_\mathrm{min}$, is 
the smallest one among all the candidate radii (see 
Fig.~\ref{fig:OptimalPlacement}(a)).

2. Disturb the position of a randomly chosen node and calculate 
the $r_\mathrm{min}$ again. If $r_\mathrm{min}$ decreases, the 
disturbance is discarded and the position of the disturbed node 
is recovered. If $r_\mathrm{min}$ is unchanged, the disturbance 
is accepted and the position of the disturbed node is updated. If 
$r_\mathrm{min}$ increases, the disturbance is accepted and the 
position of the node as well as the value of the $r_\mathrm{min}$ 
are updated accordingly.

3. Go back to step 2 and perform a new disturbance until 
$r_\mathrm{min}$ is stable.

After performing the algorithm, the final positions of the 
centers are the positions in which to place the nodes.

\acknowledgments

This work was supported by the Defense Science and Technology
Agency of Singapore under Project Agreement POD 0613356, and partially supported by the NNSF of China under Grant No. 11105025.


\begin{thebibliography}{0}


\bibitem{Newman:2003}
M. E. J. Newman, SIAM Review \textbf{45}, (2003).

\bibitem{Menezes:2004}
M. A. de Menezes and A.-L. Barab\'asi, Phys. Rev. Lett.
\textbf{92}, 028701 (2004).

\bibitem{Arenas:2001}
A. Arenas, A. D\'iaz-Guilera, and R. Guimer\`a, Phys. Rev. Lett.
\textbf{86}, 3196 (2001).

\bibitem{Guimerra:2002}
R. Guimer\`a, A. D\'iaz-Guilera, F. Vega-Redondo, A. Cabrales and
A. Arenas, Phys. Rev. Lett. \textbf{89}, 248701 (2002).

\bibitem{Toroczkai:2004}
Z. Toroczkai and K. E. Bassler, Nature, \textbf{428}, 716 (2004).

\bibitem{Tadic:2009}
B. Tadi\'c and M. Mitrovi\'c, Eur. Phys. J. B \textbf{71}, 631
(2009).

\bibitem{Xue:2010}
Y.-H. Xue, J. Wang, L. Li, D. He, and B. Hu, Phys. Rev. E
\textbf{81}, 037101 (2010).

\bibitem{Echenique:2005}
P. Echenique, J. G\'omez-Garde\~nes and Y. Moreno, Europhys. Lett.
\textbf{71}, 325 (2005).

\bibitem{Zhao:2005}
L. Zhao, Y.-C. Lai, K. Park and N. Ye, Phys. Rev. E \textbf{71},
026125 (2005).

\bibitem{Cholvi:2005}
V. Cholvi, V. Laderas, L. L\'opez, and A. Fern\'andez, Phys. Rev.
E \textbf{71}, 035103(R) (2005).

\bibitem{Korniss:2006}
G. Korniss, M. B. Hastings, K. E. Bassler, M. J. Berryman, B.
Kozma and D. Abbott, Phys. Lett. A \textbf{350}, 324 (2006).

\bibitem{Li:2010}
G. Li, S. D. S. Reis, A. A. Moreira, S. Havlin, H. E. Stanley, and
J. S. Andrade Jr., Phys. Rev. Lett. \textbf{104}, 018701 (2010).

\bibitem{Yang:2011}
H.-X. Yang, W.-X. Wang, Y.-B. Xie, Y.-C. Lai, and B.-H. Wang,
Phys. Rev. E \textbf{83}, 016102 (2011).

\bibitem{Echenique:2004}
P. Echenique, J. G\'omez-Garde\~nes and Y. Moreno, Phys. Rev. E
\textbf{70}, 056105 (2004).

\bibitem{Tang:2009}
M. Tang, Z. Liu, X. Liang, and P. M. Hui, Phys. Rev. E
\textbf{80}, 026114 (2009).

\bibitem{Yan:2006}
G. Yan, T. Zhou, B. Hu, Z.-Q Fu. and B.-H. Wang, Phys. Rev. E
\textbf{73}, 046108 (2006).

\bibitem{Danila:2006}
B. Danila, Y. Yu, S. Earl, J. A. Marsh, Z. Toroczkai and K. E. 
Bassler, Phys. Rev. E \textbf{74}, 046114 (2006); Y. Yu, B. 
Danila, J. A. Marsh, and K. E. Bassler, Europhys. Lett. 
\textbf{79}, 48004 (2007); B. Danila, Y. Sun, and K. E. Bassler, 
Phys. Rev. E \textbf{80}, 066116 (2009).

\bibitem{Zhou:2005}
S. Zhou and R. J. Mondragon, \emph{Proceedings of IEEE 
International Conference on Communications}, vol. 1 (IEEE)2005, 
pp. 163-167.

\bibitem{Krause:2006}
W. Krause, J. Scholtz and M. Greiner, Physica A \textbf{361}, 707
(2006); I. Glauche, W. Krause, R. Sollacher and M. Greiner,
Physica A \textbf{325}, 577 (2003); W. Krause, I. Glauche, R.
Sollacher and M. Greiner, Physica A \textbf{338}, 633 (2004); I.
Glauche, W. Krause, R. Sollacher and M. Greiner, Physica A
\textbf{341}, 677 (2004).

\bibitem{IEEE802.11:2007}
IEEE Std 802.11$^\mathrm{TM}$-2007, IEEE Standard for Information
technology - Telecommunications and information exchange between
systems - Local and metropolitan area networks - Specific
requirements Part 11: Wireless LAN Medium Access Control (MAC) and
Physical Layer (PHY) Specifications.

\bibitem{Moreira:2002}
A. A. Moreira, J. S. Andrade Jr., and L. A. N. Amaral, Phys. Rev. 
Lett. \textbf{89}, 268703 (2002).


\end{thebibliography}
\end{document}